\documentstyle[amssymb,aps]{revtex}

\begin{document}

\input epsf

\title{Contrasts between Equilibrium and Non-equilibrium Steady states:\\
Computer Aided Discoveries in Simple Lattice Gases}
\author{R. K. P. Zia, L. B. Shaw, B. Schmittmann, and R. J. Astalos}
\address{Center for Stochastic Processes in Science and Engineering\\
Physics Department\\
Virginia Polytechnic Institute and State University\\
Blacksburg, VA, 24061-0435 USA}
\date{\today}
\maketitle

\begin{abstract}
A century ago, the foundations of equilibrium statistical mechanics were
laid. For a system in equilibrium with a thermal bath, much is understood
through the Boltzmann factor, $e^{-{\cal H[C]}/kT}$, 
for the probability of finding
the system in any microscopic configuration ${\cal C}$. 
In contrast, apart from
some special cases, little is known about the corresponding probabilities,
if the same system is in contact with {\em more than one} reservoir of
energy, so that, even in stationary states, there is a constant energy flux 
{\em through} our system. These {\em non-equilibrium steady states} display
many surprising properties. In particular, even the simplest generalization
of the Ising model offers a wealth of unexpected phenomena. Mostly
discovered through Monte Carlo simulations, some of the novel properties are
understood while many remain unexplained. A brief review and some recent
results will be presented, highlighting the sharp contrasts between the
equilibrium Ising system and this non-equilibrium counterpart.
\end{abstract}

\vspace{1.0cm} \noindent
\nopagebreak
\vspace*{1cm}

\section{Introduction}

As we celebrate the centennial of the American Physical Society, we honor the
founding of equilibrium statistical mechanics, which also took place
about a century ago. That breakthrough enables us to understand
thermodynamics in terms of microscopics. Further, predictions based on the
Boltzmann-Gibbs framework have been applied with so much success that we now
take for granted many of the inventions of the industrial revolution,
e.g., automobiles, 747's, power stations, etc. Yet, it may be argued that no
systems are truly ``in equilibrium'', since infinite times and infinite
thermal reservoirs or perfect insulations would be necessary. Indeed,
essentially {\em all} natural phenomena bear the marks of {\em non-}%
equilibrium processes. Unlike the aforementioned class of ``artificial''
systems, most natural systems are not ``set up'' with special conditions,
under which equilibrium statistical mechanics provides excellent
approximations. Unfortunately, the theory of {\em non-equilibrium}
statistical mechanics is far less developed than its equilibrium
counterpart. As a result, the most ubiquitous phenomena are the most poorly
understood. In fact, relying on the intuition from equilibrium physics, we
are often surprised, even by the behavior of systems in {\em non-equilibrium
steady states}. These form a small subset of non-equlibrium phenomena where
the states are {\em time-independent}, mimicking systems in equilibrium. In
this article, the main differences between systems in equilibrium and
non-equilibrium steady states will be highlighted. For example, stationary
distributions of the former are well known. In contrast, we have no simple
Boltzmann-like factor, $e^{-{\cal H}/k_BT}$, for non-equilibrium steady
states. Fortunately, with the aid of modern computers, it is possible to
explore the behavior of simple model systems in stationary states far from
equilibrium. An excellent example is the driven Ising lattice gas. Despite
its simplicity, simulations continue to reveal a seemingly unending list of
counter-intuitive phenomena. Yet, because of its simplicity,
some of these surprises are now reasonably well understood.

From textbooks, we learn that the first step in equilibrium statistical
mechanics is to apply the fundamental hypothesis to a physical system in
complete isolation: every configuration, ${\cal C}$ (or microstate),
available to the system may be found with equal probability: $P_{iso}[{\cal C%
}]\propto 1$. By energy conservation, ${\cal H[C]}$ (the energy associated
with ${\cal C}$, which may include external {\em static potentials}) cannot
change, i.e., $\Delta {\cal H\equiv }0$ . Extending our scope to a system
which can exchange energy with a much larger (theoretically infinte)
reservoir and applying the fundamental hypothesis to the combination, we
arrive at the canonical ensemble: when equilibriated, the probability for
finding a system in ${\cal C}$, $P_{eq}[{\cal C}]$, is proportional to $%
\exp (-{\cal H[C]}/k_BT)$, where $T$ is the temperature associated with the
thermal reservoir. In this stationary state, on the average, the energy of
our system is a constant, which may be denoted by $\left\langle \Delta {\cal %
H}\right\rangle _T=0$. Fluctuations around this constant can be traced to
losses or gains to the reservoir, while the average energy {\em flux }%
between them is zero.

Here, we are interested in a system exchanging energies with {\em two }or
more reservoirs, which are not coupled otherwise. If, say, the reservoirs
are set at {\em different temperatures} initially, then we may expect the
following scenario. Assuming the reservoirs are much larger than our system,
then there should be a time when our system would be in a stationary state,
while the reservoirs are still close to their initial states. In this sense,
the combined system is far from equilibrium, with energy flowing from the
hotter reservoir to the colder one. However, if we focus on our system
alone, we find that its energy is constant on the average. Keeping in mind
the presence of two reservoirs (for which we use the subscripts $T$ and
$E$), we denote this situation by $\left\langle \Delta {\cal H}\right\rangle
_T+\left\langle \Delta {\cal H}\right\rangle _E=0$. This state is also far
from the equilibrium state, since there is a non-zero flux flowing {\em %
through} it. On the average, it gains a non-trivial amount of energy from
one reservoir and loses energy to the other. In other words, neither term in
the above equation vanishes: $\left\langle \Delta {\cal H}\right\rangle
_T=-\left\langle \Delta {\cal H}\right\rangle _E{\cal \neq }0$. We will
refer to such systems as being in {\em non-equilibrium steady states}.
Examples abound in nature, from our planet as a whole to simple daily
activities like cooking. The fundamental question for these states is: what
is the stationary distribution $P^{*}[{\cal C}]$?

Because our problem is inherently a time dependent one, we believe that the
most appropriate approach is to start with the $t$-dependent distribution: $%
P[{\cal C};t],$ the time evolution of which is governed by the master
equation: 
\begin{equation}
\frac \partial {\partial t}P[{\cal C};t]=\sum_{{\cal C}^{\prime }}\left\{
R\left( {\cal C}^{\prime }\rightarrow {\cal C}\right) P[{\cal C}^{\prime
};t]-R\left( {\cal C}\rightarrow {\cal C}^{\prime }\right) P[{\cal C}%
;t]\right\} \equiv {\Bbb L}P.  \label{MEq}
\end{equation}
Here $R\left( {\cal C}\rightarrow {\cal C}^{\prime }\right) $ stands for the
rate with which a configuration ${\cal C}$ changes to ${\cal C}^{\prime }$ and, in
principle, can be found once we specify how our system is coupled to the
various reservoirs. Then, the stationary $P^{*}$ will be ``just'' the (right)
eigenvector of ${\Bbb L}$ with {\em zero} eigenvalue: $0={\Bbb L}P^{*}.$

For physical systems, the task of finding the $R$'s is clearly too complex.
On the other hand, the success of equilibrium statistical mechanics suggests
those $P^{*}$'s are independent of the details of the rates. Apart from
mixing (all the configurations being connected by the $R$'s), the condition
on the rates so that our system arrives at thermal equilibrium (being
coupled to a single reservoir at temperature $T$) is known as detailed
balance: 
\begin{equation}
\frac{R\left( {\cal C}^{\prime }\rightarrow {\cal C}\right) }{R\left( {\cal C%
}\rightarrow {\cal C}^{\prime }\right) }=\exp \left\{ \frac{{\cal H[C}%
^{\prime }{\cal ]}-{\cal H[C]}}{k_BT}\right\}  \label{DB}
\end{equation}
Numerous successful Monte Carlo simulations of systems in equilibrium are
based on choosing the Metropolis rate \cite{Metropolis}: $R\left( {\cal C}%
\rightarrow {\cal C}^{\prime }\right) =\min [1,e^{-({\cal H[C}^{\prime }{\cal %
]}-{\cal H[C]})/k_BT}].$ It is clear that, with rates satisfying eqn.(\ref
{DB}), the Boltzmann $P_{eq}$ is a stationary distribution. Moreover, 
the equation $0={\Bbb L}P_{eq}$ is satisfied by 
\[
R\left( {\cal C}^{\prime }\rightarrow {\cal C}\right) P_{eq}\left( {\cal C}%
^{\prime }\right) =R\left( {\cal C}\rightarrow {\cal C}^{\prime }\right)
P_{eq}\left( {\cal C}\right) 
\]
for {\em every pair }$\left( {\cal C}^{\prime },{\cal C}\right) !$

An important distinction for a system evolving toward non-equilibrium steady
states is that eqn.(\ref{DB}) no longer holds. The effects of two reservoirs
cannot be embodied in eqn.(\ref{DB}). The stationary distribution $P^{*}$ is
not known without solving $0={\Bbb L}P^{*}$ first while {\em each term} on
the right hand side of eqn.(\ref{MEq}) is not necessarily zero. A good
analogue with electromagnetism is to regard eqn.(\ref{MEq}) as a continuity
equation. Then each ${\cal C}$ is analogous to a node in a circuit while the
terms on the right are net currents between pairs of nodes. While
equilibrium corresponds to electrostatics (with time-independent charge
distribution and zero currents), non-equilibrium steady states correspond to
magnetostatics, for which currents are steady but non-vanishing. Given this
sharp contrast between equilibrium and non-equilibrium steady states, it is
not surprising that the latter problem is considerably more difficult, since 
$P^{*}$ itself is unknown {\em a priori}. On the other hand, the variety and
richness is also much greater. In general, $P^{*}$ will depend on the
details of the rates, although we expect the ``universality classes'' of $R$%
's leading to the same $P^{*}$ to be just as large as in the equilibrium
case. In this short article, we will focus only on a particularly simple
model -- the driven Ising lattice gas \cite{KLS}. Though the model appears
simple, there is no analytic solution, so that its properties are generally
explored through computer simulations. Now, most non-equilibrium systems are
not translationally invariant. Typically, temperature gradients or velocity
shears are present, so that the usual thermodynamic limits cannot be taken.
As a result, the co-operative behavior in such systems is much more
difficult to analyse than that in systems with translational invariance. An
advantage of our simple model is that not only is it translationally
invariant, it displays a host of surprising phenomena when driven to
non-equilibrium steady states. In the next section, we will present a brief
summary of the specifications of this model and some of the remarkable
discoveries from Monte Carlo simulation studies. In many situations, the
well honed arguments of equilibrium statistical mechanics, based on the
competition between energy and entropy, fail dramatically. More details may
be found in, e.g., \cite{DL17}. Section 3 is devoted to some recent
developments while some concluding remarks are included in the last.

\section{A Brief Review of the Driven Lattice Gas}

Motivated by the physics of fast ionic conductors \cite{FIC}, Katz, Lebowitz
and Spohn introduced a simple model in 1983 \cite{KLS}, which has served as
a primary testing ground for exploring unusual properties of non-equilibrium
steady states. It consists of an Ising lattice gas \cite{Ising,YangLee} with
attractive nearest-neighbor interactions, driven far from equilibrium by an
external ``electric'' field, $E$. In the spin language, it is a
ferromagnetic Ising model with {\em biased} spin-exchange \cite
{Kawasaki} dynamics. Major reasons for choosing this model are:\newline
\indent$\cdot \ $having a translationally invariant dynamics, the steady
state is expected to be also invariant;\newline
\indent$\cdot \ $the two reservoirs are coupled through an {\em an}isotropic
dynamics;\newline
\indent$\cdot \ $many of its equilibrium properties are well-known,
especially in two dimensions ($d=2$) \cite{Onsager,McCoyWu};\newline
\indent$\cdot \ $it is a system with non-trivial phases in $d>1$, whether
driven or not;\newline
\indent$\cdot \ $the equilibrium system can be reached continuously by
taking the $E\rightarrow 0$ limit; and\newline
\indent$\cdot \ $a different equilibrium system can be reached by choosing
appropriate boundary conditions.\newline
For completeness, we give a brief description of the $d=2$ model here. On a
square lattice with fully periodic boundary conditions, each of the $%
L_x\times L_y$ sites may be occupied by a particle or left vacant, so that a
configuration ${\cal C}$ of our system is completely specified by the
occupation numbers $\{n_i\}$, where $i$ is a site label and $n$ is either 1
or 0. Translating the spin language of Ising\cite{Ising} is simple: $s\equiv
2n-1=\pm 1$. An attractive interaction between pairs of particles in
nearest-neighbor sites is modeled by the usual Hamiltonian: ${\cal H[C]}%
=-4J\sum_{<i,j>}n_in_j$, with $J>0$. The factor of 4 means that ${\cal H}$
assumes the form $-J\sum ss$ in the spin language, so that, in the
thermodynamic limit, the system undergoes a second order phase transition at
the Onsager \cite{Onsager} critical temperature $T_O=(2.2692..)J/k_B$. For
the lattice gas, this point can be reached only for half-filled systems,
i.e., $\sum_in_i=$ $L_xL_y/2$. In Monte Carlo simulations for a lattice
coupled to a thermal bath at temperature $T$, particles are allowed to hop
to vacant nearest neighbor sites with probability 
$\min [1,e^{-\Delta {\cal H}/k_BT}]$ \cite{Metropolis}, 
where $\Delta {\cal H}$ is the change in ${\cal H}$ 
after the particle-hole exchange. Note that these rules conserve the
total particle number $\sum_in_i$, so that half-filled lattices must be
used, if critical behavior is to be studied. Starting from some initial
state, this dynamics should bring the system into the equilibrium state with
stationary distribution $P_{ce}[{\cal C}]\propto e^{-{\cal H[C]}/k_BT}$.

The deceptively simple modification introduced by Katz, et. al. \cite{KLS}
is an external ``electric'' field. Pretending the particles are ``charged'',
the effect of the drive is to modify the hopping rates to biased ones.
Specifically, the jump probabilities are now 
$\min [1,e^{-(\Delta {\cal H}-\epsilon E)/k_BT}]$, 
where $\epsilon =(-1,0,1)$, for a particle attempting to hop
(against, orthogonal to, along) the drive and $E$ is the strength of the
field. Note that, locally, the effect of the external field is identical to
that due to gravity. Indeed, had we imposed ``brick wall'' boundary
conditions (particles reflected at the boundary, comparable to a floor or a
ceiling), this system would eventually settle into an {\em equilibrium}
state, like gas molecules in a room on earth. Of course, the reason behind
this outcome is that gravitation is a {\em static potential }and can be
incorporated into ${\cal H[C]}$. The ``price'' paid is the loss of
translational invariance due to the presence of boundaries, leading to
inhomogeneous particle densities.

Returning to our model, in which {\em periodic} boundary conditions are
imposed, we see that translational invariance is completely restored so
that, in the final steady state, the particle density is {\em homogeneous},
for all $T$ above some finite critical $T_c$. At the same time, a particle
current will be present. For gravity, such a situation exists only in art 
\cite{MCE}. In physics, this situation can be realised only with an {\em %
electric} field. Thanks to Faraday, a constant electric field circulating
around the surface of a cylinder can be set up by applying a linearly
increasing, axial magnetic field. If the particles are charged, they will
experience the same force everywhere on the cylinder. As they also lose
energy to the thermal bath, a steady state with finite current can be
established. With this possibility in mind, we will use the term
``electric'' field to describe the external drive and imagine our particles
to be ``charged''. We should caution the reader that, unlike real charges,
our particles are not endowed with Coulomb interactions between them, just
like the typical neglect of gravitational attraction between gas molecules.
Finally, note that a well defined, single-valued, {\em potential} which
gives rise to such an electric field is necessarily time dependent.

Given the microscopic model, we can ask: What are its collective properties
when it settles down in a non-equilibrium steady state? In particular, what
is $P^{*}[{\cal C}]$? Since there is no global Hamiltonian, we cannot
exploit Boltzmann's result. Instead, we must resort to the master equation (%
\ref{MEq}) and attempt to find the solution to $0={\Bbb L}P^{*}.$ Though $%
{\Bbb L}$ is a sparse matrix, finding $P^{*}$ is a non-trivial task. Only
for very small systems can $P^{*}$ be obtained \cite{Small}. Of course, it
is difficult to discern collective behavior such as phase transitions in
``microscopic'' systems like these. Nevertheless, we can already see that
the $P^{*}$'s here (fig. 1a) are quite distinct from $P_{eq}$ (fig. 1b).
Also, at this level, we can derive an interesting consequence: the violation
of the standard fluctuation dissipation theorem. Computing $\left\langle 
{\cal H}\right\rangle $ and $\left\langle {\cal H}^{2}\right\rangle $, it is
easy to verify that, in the driven case, $\left\langle {\cal H}%
^{2}\right\rangle -\left\langle {\cal H}\right\rangle ^{2}\neq -\partial
_{\beta }\left\langle {\cal H}\right\rangle .$%

Turning to collective phenomena on the macroscopic scale, we face serious
difficulties in finding $P^{*}$ analytically, let alone solving for
thermodynamic quantities a la Onsager. Without modern computers, it would be
impossible to make much progress. Using simulation techniques, we may answer
questions like: what happens to the second order phase transition found by
Onsager? In particular, how does the critical temperature depend on the
drive, i.e., what is $T_c(E)$? Several simple possibilities come to mind:%
\newline
\indent a) $T_c(E)$ jumps to infinity for any $E$ , i.e., the drive, however
small, orders the system. \newline
\indent b) $T_c(E)$ rises with $E$ indefinitely. \newline
\indent c) $T_c(E)$ rises with $E$, saturating at some finite temperature $%
T_c(\infty )$. \newline
\indent d) $T_c(E)$ is independent of $E$ , i.e., $T_c(E)=T_O$. \newline
\indent e) $T_c(E)$ decreases with $E$, either dropping to zero at finite $E$%
, or saturating at a finite $T_c(\infty )$. \newline
\indent f) $T_c(E)$ jumps to zero for any $E$ , i.e., the drive, however
small, disorders the system. \newline
While possibilities (a), (b), and (f) appear incredulous, intuitive
arguments might be made for (d) and (e). The naive argument for (d) is that,
in an inertial frame where the global current vanishes, the system should
look just like an equilibrium Ising model. On the other hand, to arrive at
(e), we think of the drive as gravity, feeding energy into the system, so
that its effects should be the same as a reservoir with a temperature higher
than the surrounding bath. Therefore, to order the system, the bath
temperature would have to lowered, i.e., $T_c(E)<T_O$. In reality,
simulations \cite{KLS} offered the first surprise: $T_c(E)$ {\em increasing}
with $E$ and saturating at $T_c(\infty )\simeq 1.4T_O$ \cite{KTLJSW}! A
number of arguments now exists for this behavior, but all are
``post-dictions''. Indeed, none of these are convincing, while approximate
schemes for analytic computations of $T_c(E)$ offer only hints to the puzzle
of $T_c(E)>T_c(0)$ \cite{YS,HTS}.

As the system is probed deeper, more surprises appear. Due to space
limitations, we only list some of them here, refering the interested reader
to \cite{DL17,ALB} for further details.


\begin{figure}
\vspace*{-0.5cm}
\hspace*{-1cm} 
\epsfxsize=8in \epsfbox{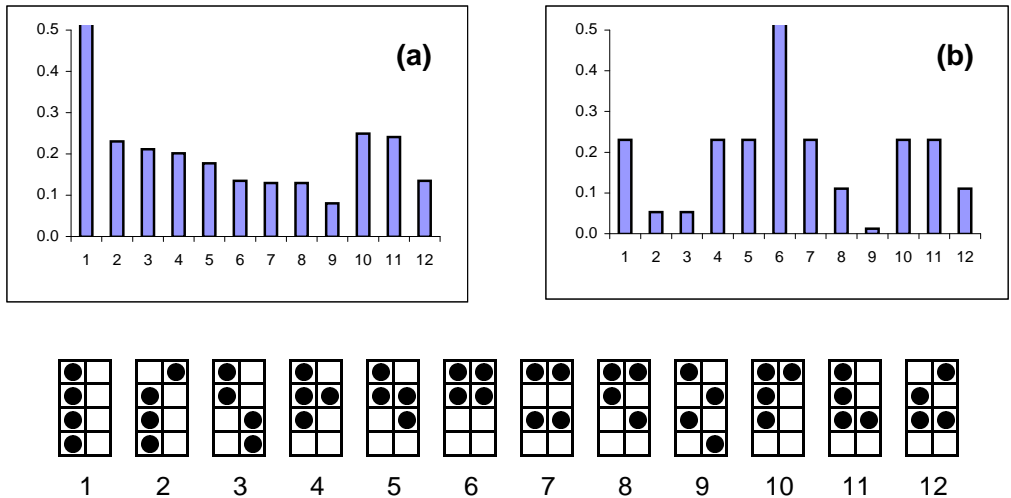}
\vspace*{-18.5cm}
\caption{Stationary distributions for a 4$\times $2 Ising lattice gas,
driven with infinite $E$ (a) and in equilibrium (b).
The normalization is set via: largest $P^{*}\equiv 1$.
Corresponding configurations (only one out of the equivalence class) are
shown, with the drive downwards. }
\end{figure}


\subsection{Disordered phase ($T>T_c$)}

In the equilibrium system, there is little of interest far above
criticality. Correlations are short ranged so that most properties can be
understood through Landau-Ginzburg mean field theory. When driven, however,
this system displays\newline
\indent $\cdot \ $long range two-point correlations, decaying as $1/r^d$ 
\cite{LR}\newline
\indent $\cdot \ $singular structure factors, with a discontinuity at the
origin \cite{KLS,SingSF}\newline
\indent $\cdot \ $non-trivial three point correlations, the Fourier
transforms of which show infinite discontinuities at the origin \cite{3pf};%
\newline
\indent $\cdot \ $a fixed line, rather than the Gaussian fixed point,
governing large-scale, long time behavior \cite{fixedline};\newline
\indent $\cdot \ $shape-dependent thermodynamics\cite{Eyink}

\subsection{Critical Behavior ($T\sim T_c$)}

In 1944, Onsager solved the $d=2$ Ising model and computed many of its
critical properties. However, the deeper understanding of critical phenomena
came only in the 70's, with the advent of field theoretic renormalization
group analysis\cite{RG}. Within this framework, we learnt that a large class
of systems fall into the Ising universality class, controlled by the
Wilson-Fisher fixed point\cite{WF}. When driven into non-equilibrium states, 
\newline
\indent $\cdot \ $only ordering into strips parallel to $E$ occurs;\newline
\indent $\cdot \ $only one of the two lowest structure factors diverges;%
\newline
\indent $\cdot \ $strong anisotropy appears while longitudinal and
transverse momenta scale differently;\newline
\indent $\cdot \ $the critical dimension is 5 instead of 4; \newline
\indent $\cdot \ $a new, non-Hamiltonian, fixed point and universality class
is identified \cite{JSLC};\newline
\indent $\cdot \ $a host of new exponents, though only one independent,
emerges; and\newline
\indent $\cdot \ $anisotropic finite size scaling is essential for data
collapse \cite{KTLJSW}. \newline
We should note that, unlike properties far from $T_c$, the critical
properties were predicted by theory \cite{JSLC} well before confirmations
from computer simulation studies \cite{KTLJSW}.

\subsection{Ordered phase ($T<T_c$)}

Below the critical point, phase segregation and co-existence occurs, with
interfaces separating the particle rich from the particle poor regions. Due
to the periodic boundary conditions, each region is a single strip wrapped
around the torus. Unlike in equilibrium, only strips {\em parallel} to the
drive exist. Correlations within each region are also expected to be long
ranged, though they have not been studied carefully so far. Some intriguing
properties of the interfaces, dramatically different from equilibrium \cite
{eqInterface} ones, are: \newline
\indent $\cdot \ $statistical widths remaining finite as $L\rightarrow \infty 
$ \cite{LMVZ} rather than diverging as $\sqrt{L}$; \newline
\indent $\cdot \ $the structure factor diverging as $q^{-0.67}$ \cite{LZ93}
instead of the usual $q^{-2}$ of capillary waves\cite{cap-wav};
\indent $\cdot \ $interface
orientation affecting {\em bulk} energies \cite{SPBC}; and\newline
\indent $\cdot \ $instabilities when forced to lie at a non-vanishing angle
with respect to the drive\cite{L88Y93}. \newline
In addition to interface anomalies, other remarkable features include\newline
\indent $\cdot \ $coarsening during a quench showing several time regimes
and asymmetries which cannot be accounted for by a modified Cahn-Hilliard theory 
\cite{CH};\newline
\indent $\cdot \ $systems subjected to {\em shifted} periodic boundary
conditions displaying new, multistrip phases \cite{SPBC}; and\newline
\indent $\cdot \ $systems subjected to {\em open} periodic boundary
conditions displaying icicle like fingers\cite{OBC}.

As we try to convey in this section, there are still many unresolved
mysteries associated with this decepetively simple model. In the next
section, we will present some recent, on-going investigations which attempt
to probe deeper into the simple driven lattice gas.

\section{Some Recent Developments}

\subsection{Steady State Energy Fluxes}

In the introduction, we emphasized the importance of energy flow {\em through%
} our system in a non-equilibrium steady state. There has been no systematic
study of this aspect, even though it may contain a key to the understanding
of such systems. We begin by simply confirming our intuitive picture: that
typical jumps parallel/trasverse to the field are associated with energy
gain/loss. In the previous section, we showed the complete solution for a $%
4\times 2$ lattice. Using those $P^{*}$'s, we can compute these gains/losses
and then compare them with Monte Carlo simulations \cite{RJA}. The results
are best displayed as histograms for all possible values of $\Delta {\cal H}$
after an attempted jump. There are two such histograms, associated with the
two types of jumps. We also performed the same analysis for the equilibrium
case. It is reassuring that simulations confirm, within statistical errors,
all theoretical predictions. Encouraged by these results, we carried out
simulations on a $30\times 30$ lattice \cite{RJA}. Figure 2a shows that, for
the equilibrium case, both histograms are entirely symmetric (within
statistical errors), so that the gains/losses balance for either types of
jumps. By contrast, figure 2b shows asymmetric histograms, confirming our
expectation that the systems tends to gain energy when a particle jumps in
the field direction, etc. Our hope is that a good approximation scheme,
within the theoretical framework described above, can be found leading to
quantitative predictions of these histograms.

\subsection{A Possible New Phase in Lattice Gases with Anisotropic
Interactions}

Since the drive introduces a non-trivial anisotropy into the Ising lattice
gas, it is natural to ask how $E$ might compete with anisotropic {\em %
couplings}. The equilibrium model is part of Onsager's solution, so that we
may again compare driven cases with well-known results. Since the drive
enhances longitudinal correlations, we had expected that ``$T_c(E)$ will be
higher (lower) if the drive is aligned with the stronger (weaker) bonds''
\cite{Bilayer}. Subsequent simulations showed the opposite\cite{LBS}!
Indeed, with saturation drive, $T_c(E)$ can drop {\em below} $T_c(0)$ for 
$\alpha \gtrsim 1.7$, where $\alpha ^2$ is the ratio of the coupling along
the field direction to the ``transverse'' coupling. Motivated to look into
the behavior displayed at various temperatures, we find that the typical
configurations are indeed disordered for $T\gg T_c(0)$ and ordered into a
single strip for $T\ll T_c(E)$. However, there is a significant range of
temperatures where the system appears to be ``ordered'' in the drive
direction without being in a single strip. In other words, while the
densities in each column (i.e., along the drive) are bi-modally distributed 
\cite{LBSMS}, the usual order parameter ($S(1,0)$, structure factor with
lowest transverse wave number) is still quite small. In more picturesque
language, we call this a ``stringy'' state. Actually, this type of
configurations have been previously reported \cite{Marro}. But they were
believed to be long transients on the way from disordered initial states to
ordered ones and thus, disregarded. By contrast, we observed that, starting
from ordered states, the system evolves towards, and spends considerable
periods of time in, the stringy states.


\begin{figure}
\vspace*{-0.5cm}
\epsfxsize=7in \epsfbox{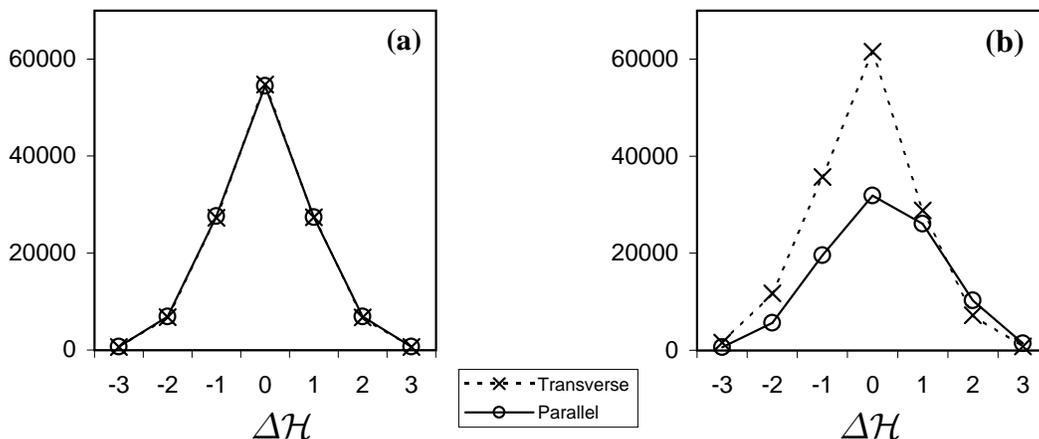}
\vspace*{-16.5cm}
\caption{Histograms of $\Delta {\cal H}$ after an attempted jump in a 
$30\times 30$ lattice, at $T=2.5$ with $E=0$ (a) and $E=\infty $ (b).}
\end{figure}


In an effort to quantify these states, we define the ratio: 
\begin{equation}
{\cal R}\equiv L^d\frac{G(0,L/2)}{S(1,0)}  \label{R}
\end{equation}
where $G(x,y)$ is the (untruncated!) two point correlation function, with $y$
being parallel to the drive, and $S(k,p)\equiv \sum_{x.y}G(x,y)e^{2\pi i\left(
kx+py\right) /L}$ is the structure factor, used ordinarily as the order
parameter. In particular, we are interested in the behavior of ${\cal R}$ as 
$L\rightarrow \infty $.

Far in the disordered phase, as a result of the $r^{-d}$ decay, we have $%
G(0,L/2)\rightarrow O(L^{-d})$ while $S(1,0)\rightarrow O(1)$. On the other
hand, deep in the ordered phase $G\rightarrow O(1)$ while $S\rightarrow
O(L^d)$. Therefore, as long as the system is far from criticality, we have

\begin{equation}
{\cal R}\rightarrow O(1).  \label{far-from Tc}
\end{equation}
Near criticality, if there are no stringy states, we may apply the results
of renormalization group analysis \cite{JSLC} -- $G(0,r)\rightarrow
r^{-[d-2+\Delta ]/(1+\Delta )}$ and $S(k,0)\rightarrow k^{-2}$ -- to arrive
at ${\cal R}\rightarrow L^{\left[ d-3\right] \Delta /(1+\Delta )}$. Here, $%
\Delta $ is the exponent associated with anisotropic scaling of the momenta (%
$k_{\Vert }\sim k_{\bot }^{1+\Delta }$), and is found to be $(8-d)/3$ to all
orders in the expansion around $(5-d\dot{)}>0.$ Thus, for our simulations in 
$d=2$, we see that ${\cal R}$ {\em decreases} with $L:$

\begin{equation}
{\cal R}\rightarrow L^{-2/3}  \label{DDS-Tc}
\end{equation}
By contrast, {\em in a stringy state}, ordering has set in for the
longitudinal direction so that $G(0,L/2)\rightarrow O(1)$. Meanwhile,
complete phase segregation into just two regions is yet to take place, so
that $S(1,0)\rightarrow O(1)$. As a result, we expect

\begin{equation}
{\cal R}\rightarrow L^d  \label{stringy}
\end{equation}
which is an {\em increasing} function of $L.$ The differences between the
three behaviors (\ref{far-from Tc}, \ref{DDS-Tc}, and \ref{stringy}) should
be dramatic enough to discern in simulations.

To provide further contrast, we perform the above analysis for Ising models
in {\em equilibrium}. Above criticality, $G\rightarrow e^{-r/\xi }r^{2-d}$
and $S\rightarrow O(1)$ so that ${\cal R}$ decreases exponentially as $%
L^2e^{-L/\xi }$. Near $T_c,$ we have $G\rightarrow r^{2-d+\eta }$ and $%
S\rightarrow k^{-2+\eta }$ leading to ${\cal R}\rightarrow O(1)$. The
behavior deep in the ordered phase is unchanged. So, in all cases, ${\cal R}$
should {\em not increase with }$L$.

Turning to simulations, we find that the stringy state seems to be most
pronounced for large $\alpha .$ In particular, we focused on models with 
$\alpha =3$ and saturation drive and compiled data from systems sizes 
$L=10,20,30,40,60,$ and $90$, using two or three independent runs. Starting
with both random and ordered initial conditions, the runs last up to 800,000
Monte Carlo steps. The plot of ${\cal R}$ vs. $L^2$ in figure 3 shows that,
for two of the temperatures investigated, this ratio appears to be {\em %
increasing with }$L$. Meanwhile, for $T\notin \left[ 0.8,1.0\right] $, the
behavior is consistent with ${\cal R}\rightarrow O(1)$. These preliminary
results lead us to conjecture the existence of a ``stringy phase'',
especially in the thermodynamic limit of square ($L\times L$) lattices. Of
course, simulations with larger $L$'s will be needed to determine if the
increasing behavior persists. Other systematic methods can also be brought
to bear, such as distribution functions of both $G$ and $S$. We have
initiated a study of the histograms of $S(k,p)$ for a range of low lying
wave-vectors, using the time-series of each quantity.
Preliminary data show remarkable structures which are being verified in
longer runs. Attempts at a theoretical understanding, based on
phenomenological approaches, of the nature of the ``stringy phase'' are in
progress.


\begin{figure} [htbp]
\epsfxsize=7in \epsfbox{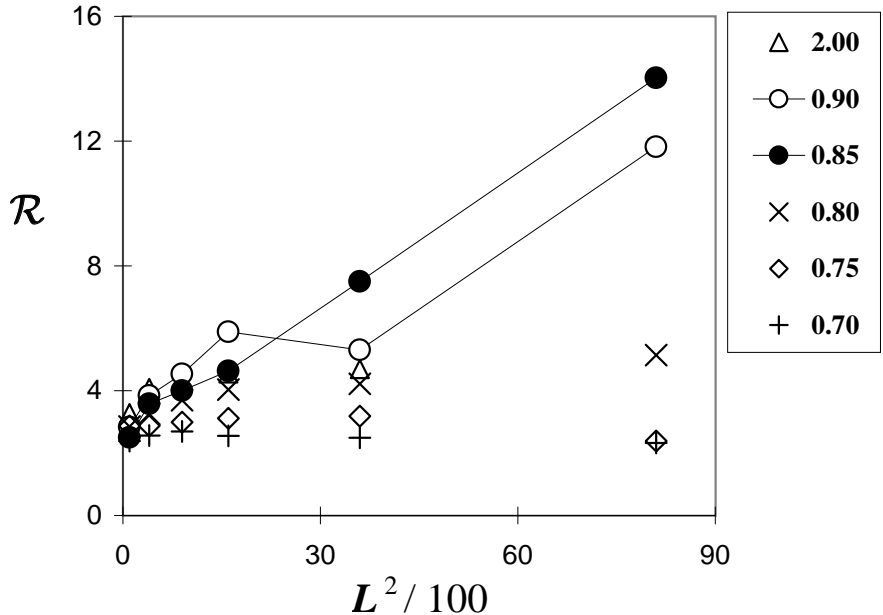}
\vspace*{-14cm}
\caption{Correlation-Structure Factor ratio ${\cal R}$ for systems with $%
\alpha =3$ and saturation drive, plotted against $L^2.$ The legend refers to
the temperatures of the various runs.}
\end{figure}


\section{Concluding Remarks}

In this brief article, we highlighted several major differences between a
system in thermal equilibrium and one in non-equilibrium steady states.
Apart from the obvious presence of non-trivial energy fluxes, the latter
systems display many distinguishing and surprising features. Since their
stationary distributions are neither {\em a priori} known nor susceptible to
analytic probes (except for some simple limits or 1-D cases), all efforts to
uncover the macroscopic, collective properties of these systems are based on
Monte Carlo simulations.

Focusing on a particularly simple model -- the Ising lattice gas, driven
into non-equilibrium steady states by an external ``electric'' field\cite
{KLS}, we gave a brief review of a variety of surprising and
counter-intuitive behavior. In the last section, we offered two of the recent
developments in this continuing saga: detailed investigations of the energy
flux and preliminary studies indicating the possible existence of a new
phase (especially for systems with large anisotropies in the attractive
interactions). Beyond this simple model, many generalizations have been
explored. Examples include repulsive interactions, random drives, quenched
random impurities, multilayers, and multispecies, to name but a few \cite
{DL17}. Further from this class of ``driven diffusive systems'' is a wide
range of other non-equilibrium steady states, e.g., surface growth,
electrophoresis and sedimentation, granular and traffic flow, biological and
geological systems, etc. All of these offer further surprises, some
understood and most unexplained. At present, each non-equilibrium
system is studied independently from the others. The hope is that a unifying
concept and framework, like the fundamental hypothesis or the Boltzmann
factor, will be discovered before the next centennial meeting of the APS.

\section{Acknowledgments}

We are grateful to F.S. Lee for some unpublished work on the 4$\times $2
system. This research was supported in part by a grant\ from the National
Science Foundation through the Division of Materials Research.

\end{document}